# Superplasticity-like behavior of ultrafine-grained austenitic steel 321

V.I. Kopylov[1], M.Yu. Gryaznov[1], S.V. Shotin[1], A.V. Nokhrin[1,(*)], C.V. Likhnitskii[1], M.K. Chegurov[1], V.N. Chuvil'deev[1], N.Yu. Tabachkova[2,3]

[1] Lobachevsky State University of Nizhniy Novgorod, Nizhniy Novgorod, Russia

[2] National University of Science and Technology "MISIS", Moscow, Russia

[3] A.M. Prokhorov Institute of General Physics, Russian Academy of Science, Moscow, Russia

nokhrin@nifti.unn.ru

**Abstract**

Hot rolled commercial metastable austenitic steel 321 with strongly elongated thin δ-ferrite particles in its microstructure was the object of investigations. Ultrafine-grained (UFG) microstructure in steel 321 was formed by Equal Channel Angular Pressing (ECAP) at 150 ºC and 450 ºC. When heating the UFG steel specimens, the nucleation of σ-phase particles blocking the grain boundary migration was observed. The maximum elongation to failure (~250%) was achieved at the deformation temperature 750 ºC. The process of superplastic deformation of the UFG steel 321 is controlled by simultaneous grain boundary sliding and power-law creep. The contribution of each process depends on the grain growth rate in the superplasticity regime as well as on defect accumulation on the grain boundaries. The fracture of the UFG steel 321 specimens has a cavitational character – an intensive formation of large elongated pores at the non-metallic particles as well as of the submicron pores at the σ-phase particles in the course of superplastic deformation were observed.

---

(*) Corresponding author (nokhrin@nifti.unn.ru)

**Introduction**

The stainless austenitic steels (304, 316, 321, etc.) are applied extensively in the petrochemical industry, in nuclear power engineering, and in mechanical engineering. Low strength of coarse-grained (CG) austenitic steels is one of the drawbacks of these ones, which doesn't allow providing necessary level of reliability of the products.

The formation of the ultrafine-grained (UFG) microstructure in the austenite steels is a conventional method of improving the strength of these ones. The deformation of the austenite steels at room or moderate temperatures followed by annealing is the most common method of forming the UFG microstructure [1-3]. The cold plastic deformation of the austenitic steel leads to the strain-induced martensite formation while annealing initiates the reverse transformation of martensite into austenite [2-5]. Severe Plastic Deformation (SPD) at elevated temperatures allows forming UFG austenite microstructure without additional annealing [6-8]. UFG austenitic steels obtained by SPD have high characteristics of strength and hardness as well as satisfactory ductility [9-18], and good corrosion resistance [19-24].

The ductility of the CG steels is low even at elevated temperatures, and the relative elongation to failure in the tension tests exceeds 50-100% seldom [1]. The forming of the UFG microstructure allow increasing the ductility of steels at elevated temperatures and small strain rates [25-28]. Earlier, it was shown that the UFG materials can demonstrate very high ductility at high strain rates and reduced temperatures [29-32]. The decreasing of the temperature and of the deformation time at simultaneous increasing of the ductility allows reducing the power consumption of the enterprises, replacing the die tooling for hot stamping, making the products of more complex shapes with the minimal damage, decreasing the oxidation of the steel, etc.

Note that in spite of a great interest to the superplasticity of ferritic and duplex steel [33], there are very few works on the superplasticity of the austenitic steels obtained by SPD [34-39]. At present, major attention is attracted to the superplasticity of steels, the fine-grained microstructure in which is formed by the reverse transformation of martensite into austenite when heating [34-37, 39].

The low-temperature superplasticity steel 304, the UFG microstructure in which was formed by cold rolling followed by annealing was described in [34]. After annealing at 700 °C for 5 min, the steel had completely austenite UFG microstructure. At the room temperature, the σ(ε) curves had the plastic instability region, which was typical for the Lüders deformation. At 600-650 °C, the σ(ε) curves had the shapes typical for the superplastic flow. For the UFG steel 304, the maximum elongation of 268-296% was achieved at the 600 °C and the strain rate of $2.5 \cdot 10^{-4}$ s$^{-1}$. The strain rate sensitivity coefficient was 0.22-0.36. The samples with completely UFG austenite microstructure demonstrated al lower ductility. An intermittent flow was observed in the σ(ε) curves for the UFG steel 304 with the maximum ductility. Since the austenite grains remained equiaxial after testing, the authors of [34] concluded the grain boundary sliding (GBS). Note also that the formation of elongated pores of several microns in sizes was observed at the superplasticity of the UFG steel 304.

The authors of [34, 36] suggested residual martensite to suppress the dynamic austenite grain growth at the superplasticity of UFG steel. Note the work [36] where the elongation to failure of UFG steel 18Cr-9Ni was shown to depend on α′-martensite content strongly. The maximum ductility in the UFG steel 18Cr-9Ni was achieved at the α′-martensite content of ~10% - the elongation to failure was 270% at 650 °C and the strain rate $10^{-3}$ s$^{-1}$ [36].

In [35], a comparative analysis of mechanisms of deformation for CG and UFG steel 18Cr-8Ni was carried out. The CG and UFG steels had the mean austenite grain size was 9 μm and 1.5 μm respectively. The maximum elongation to failure for the CG steel was 113% at 800 °C. In the temperature range 25-400 °C, the ductility of UFG steel was smaller than the one of the CG steel whereas at 600-800 °C it was close to the ductility of the CG steel or slightly higher. The curve σ(ε) for the UFG steel at 800 °C had an elongated stage of stable plastic flow whereas the σ(ε) curves at 25-600 °C were featured by an intensive strain hardening. The authors of [35] didn't observe the effect of intermittent plastic flow of the samples of CG and UFG steel at room and elevated temperatures. It is interesting to note that higher martensite content was observed in the samples of UFG steel in the tension tests. The authors of [35] related the presence of a large number of small recrystallized grains at the grin boundaries after straining at 600 and 800 ºC to GBS and the beginning of the dynamic recrystallization.

In [37], an achievement of very high ductility for the UFG steel 304 was reported. The UFG steel had a grain sizes of ~0.2 μm and martensite content was small (< 0.5%). At 600 ºC and the strain rate < $3 \cdot 10^{-4}$ s$^{-1}$, the elongation to failure was > 500%. The magnitude of the coefficient *m* decreased from ~0.5 down to 0.28-0.30 with increasing strain rate from $10^{-3}$ s$^{-1}$ up to $1.8 \cdot 10^{-2}$ s$^{-1}$ at 600 ºC. The authors of [37] referring to their earlier works, also reported on the possibility to achieve the elongation to failure in steel 304 > 600% at the ~$5.5 \cdot 10^{-5}$ s$^{-1}$ and 600 ºC.

In [39], unexpected results of investigations of superplasticity of fine-grained austenite steels were presented. The ductility of CG metastable austenite steels is known to depend on the heating temperature nonmonotonously [34, 40, 41]. The maximum elongation to failure ($\delta_{max}$ = 43%) of the fine-grained steel 321L was achieved at 500 ºC. This value exceeds the ductility of the CG steel at this temperature (~6%). Among other interesting results presented in [39], one should outline the absence of the ductility reduction effect in the temperature range 400-500 ºC in the UFG steel 321.

Summarizing the results of brief analysis, one can conclude that there are some contradictions in the question of the mechanisms of superplasticity of UFG steels as well as in the question of the effect of various factors on the strain character in the UFG austenitic steels in the superplasticity conditions at present.

The present study was aimed at the investigation of the superplastic deformation mechanisms in the UFG austenitic steel 321L at the temperatures lower than 900 ºC. This temperature corresponds to the beginning of dynamic recrystallization at conventional strain rates [6, 42]. To date the superplastic deformation of this steel in the UFG state hasn't been studied yet. The choice of steel 321L as the object of investigations was motivated by its wide application in oil chemical industry, in mechanical engineering, and in nuclear power engineering. The heat exchangers, muffles, pipes, shut-off valve parts, and other parts operated in the conditions of action of elevated temperatures and corrosion-aggressive media are fabricated from Ti-containing steel 321L by hot deformation.

**Materials and methods**

The commercial metastable austenitic steel 321L (composition, in wt.%: Fe-0.08C-17.9Cr-10.6Ni-0.5Si-0.1Ti, Russian trademark 08X18H10T) was the object of investigations. The formation of the UFG microstructure in the steel was performed by ECAP. The workpieces of 14×14×70 mm in sizes were cut out from hot-rolled rods of 20 mm in diameter. Prior to ECAP, the rods were annealed at 1050 ºC for 30 min followed by quenching in water. ECAP was performed using Ficep® HF400L press (Italy). The angle of crossing the working channel and the output one was 90º ($\pi/2$). The workpieces were processed in "A" ECAP route. The ECAP temperatures were 150 and 450 °C, the number of pressing cycles (N) varied from one to four. After every ECAP cycle, the workpieces were cooled down the room temperature, greased, installed into the work channel of the ECAP punch, and heated up during 15 min prior to the next ECAP cycle. Graphite grease with addition of $MoS_2$ was used in ECAP. The ECAP rate was 0.4 mm/s.

The investigations of the steel microstructure were carried out using Jeol® JSM-6490 and Tescan® Vega™ Scanning Electron Microscopes (SEMs) and Jeol® JEM-2100F Transmission Electron Microscope (TEM). X-ray diffraction (XRD) phase analysis of the steels was carried out using Shimadzu® XRD-7000 X-ray diffractometer ($CuK_\alpha$ emission, scanning step – 0.02°, exposure time – 3 s). The mean grain size (d) was calculated by section method using GoogGrains 2.0 software. The local chemical analysis of microstructure were studied using Oxford Instruments® INCA 350 EDS spectrometer. The unit cell parameters and mass fraction of martensite were calculated by Rietveld method using Topas software. The phase identification was performed using PDF and ISCD databases.

The microhardness (Hv) of the steel was measured with Duramin® Struers™ 5 microhardness tester. The measurements of Hv were performed at the load of 2 kg. The uncertainty of the microhardness measurements was ± 50 MPa. For the mechanical tests, flat double-blade shaped (dogbone-shaped) specimens. The sizes of the working part of the dogbone-shaped specimens were 2×2×3 mm (Fig. 1). The tension tests were carried out using Tinius Olsen® H25K-S machine with the strain rates $3.3 \cdot 10^{-4}$ s$^{-1}$ (the tension rate was $10^{-3}$ mm/s) and $3.3 \cdot 10^{-3}$ s$^{-1}$ (the tension rate was $10^{-2}$ mm/s). The tension tests were performed at the room temperature (RT) and in the temperature range 450–900 °C. The specimens were heated up to the testing temperatures in 5 min. The specimens were kept at the testing temperatures for 10 min to establish the thermal equilibrium. From curves stress ($\sigma$) – strain ($\varepsilon$), the magnitudes of the ultimate strength ($\sigma_B$) and of the relative elongation to failure ($\delta$) were determined.

The fractographic analysis of the fractures after the tension tests was carried out using Jeol® JSM-6490 SEM. The main zones on the fracture surface (the fibrous, radial, and cut zones) were distinguished by comparing the SEM results with [43]. The macrostructure of the specimens after the failure tests was investigated using Jeol® JSM-6490 SEM and Leica® IM DRM metallographic optical microscope. The investigations of microstructure and the microhardness measurements were performed in the fracture zones ("deformed area") and in the non-deformed areas near the capturers (Fig. 1).

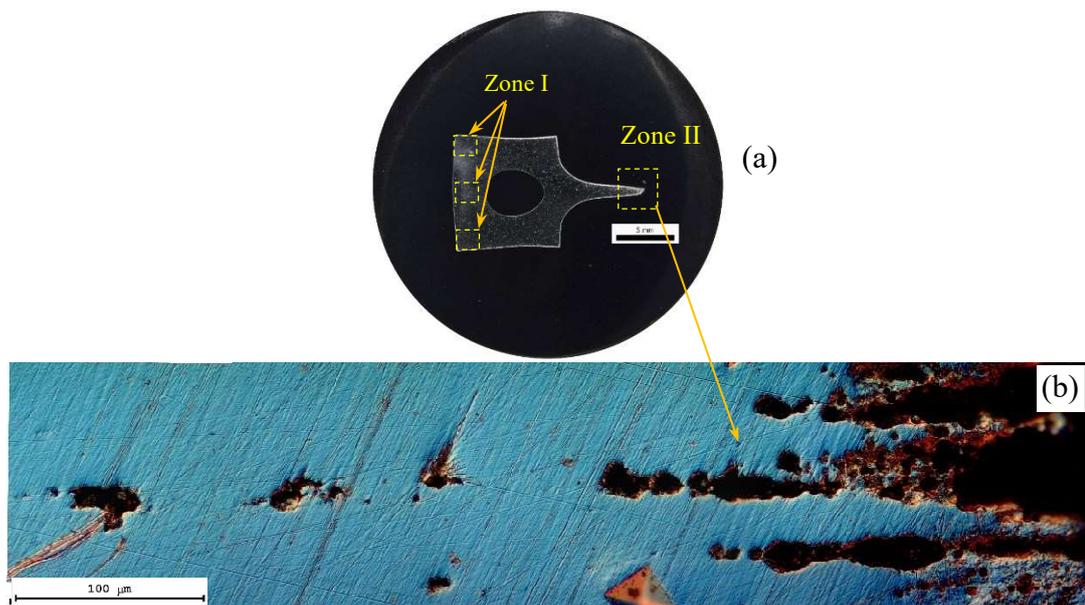

Figure 1. General appearance of UFG steel sample (ECAP, N = 3, 150 °C) after tension testing at 800 °C (a). In Fig. 1a, the areas of microstructure investigations and microhardness measurements are marked: Zone I is the non-deformed one, Zone II is the deformed one (the fracture zone). In Fig. 1b, an enlarged image of the fracture zone is presented

**Results**

<u>Microstructure investigations</u>

Metastable steel 321 in the initial state had a uniform CG austenite microstructure (Fig. 2a), which thin δ-ferrite particles were present in. The thicknesses of the δ-ferrite particles were ~3-5 μm (Fig. 2a), the lengths of the ones exceeded 500 μm. There were some non-metallic particles in the steel microstructure, which may form the agglomerates (Fig. 2b) or be located separately. The non-metallic particles comprised mainly TiN and Ti(C,N) particles. In Fig. 2a, the largest separate particles are marked by white lines. The XRD phase analysis results revealed the mass fraction of δ-ferrite in the CG steel to be ~2% (Fig. 3).

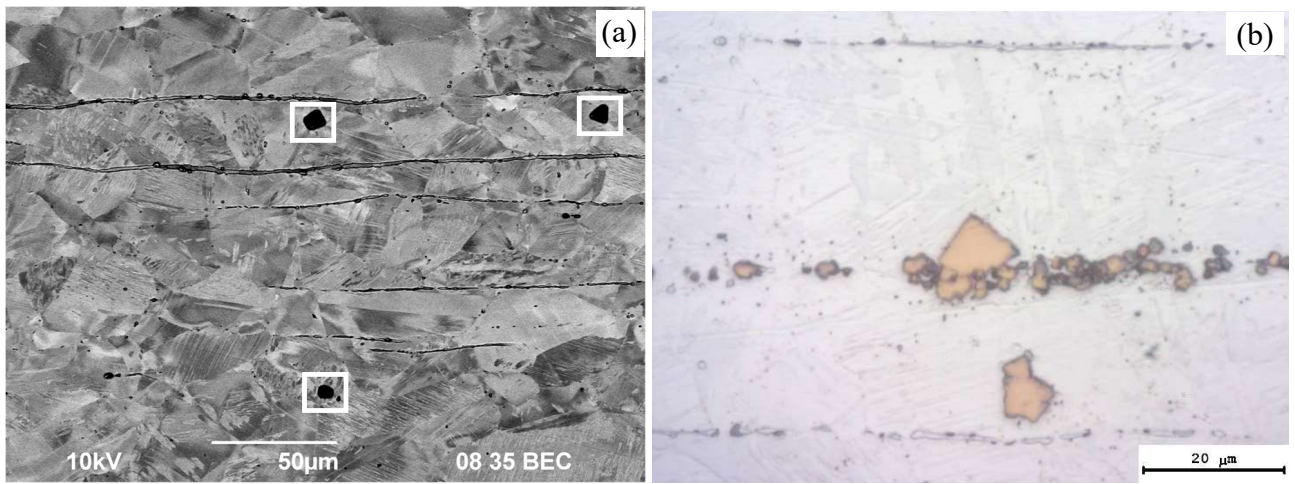

Figure 2. Microstructure of CG steel (a) and large particles (b) in the CG steel: (b) line-by-line arrangement of large particles

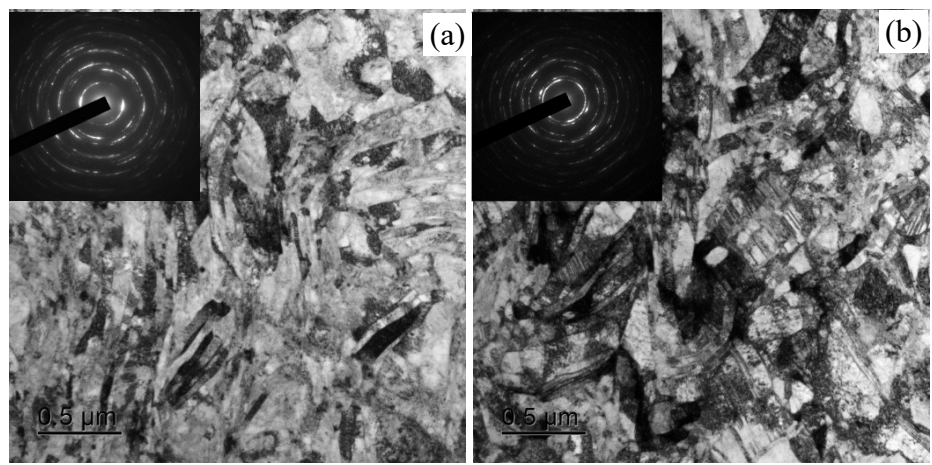

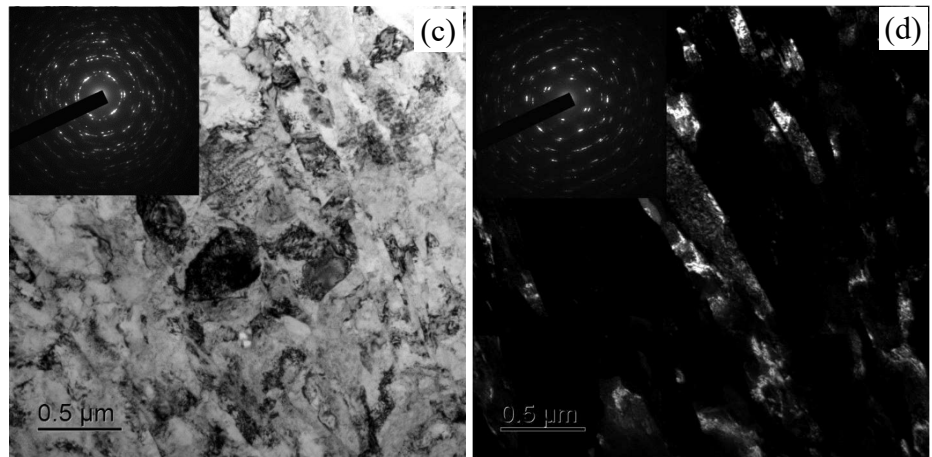

Figure 3. Microstructure of UFG steel after ECAP (N = 4) at 150 °C (a, b) and 450 °C (c, d). TEM [44]

The ECAP process led to the formation of the UFG austenite microstructure and to strain-induced formation of α′-martensite. As one can see in Fig. 2c, austenitic steel 321 after ECAP at 150 °C had a strongly fragmented austenite microstructure. The microstructure of the samples was studied in the cross section of the UFG workpiece. The mean sizes of the austenite fragments were 0.3-0.5 μm. The deformation twins are seen inside the grains (Fig. 2d). The grains in the microstructure of the UFG steel after ECAP at 450 °C had more equiaxial shapes (Fig. 2e, f). The mean austenite grain sizes after N = 4 ECAP cycles at 150 and 450 °C were 0.3-0.4 μm and 0.5-0.7 μm, respectively. No δ-ferrite strips were found in the microstructure of the UFG steel after ECAP. The sizes and the character of spatial distribution of the large non-metallic particles in the UFG steel samples were close to respective parameters of the CG steel. Note that even after ECAP at 450 °C steel had two-phase austenite-martensite microstructure. As one can see in Fig. 3, the (110) α′-martensite peaks are present in the XRD curves of the samples of UFG steel. The mass fractions of α′-martensite after four ECAP cycles at 150 °C and 450 °C were ~5-15%. The α′-martensite content didn't depend on the ECAP temperature within the measurement uncertainty that corresponds to the results of [45]. No peaks corresponding to ε-martensite (PDF #00-034-0529, ICSD #631723) were observed in the XRD curves of the UFG steel samples.

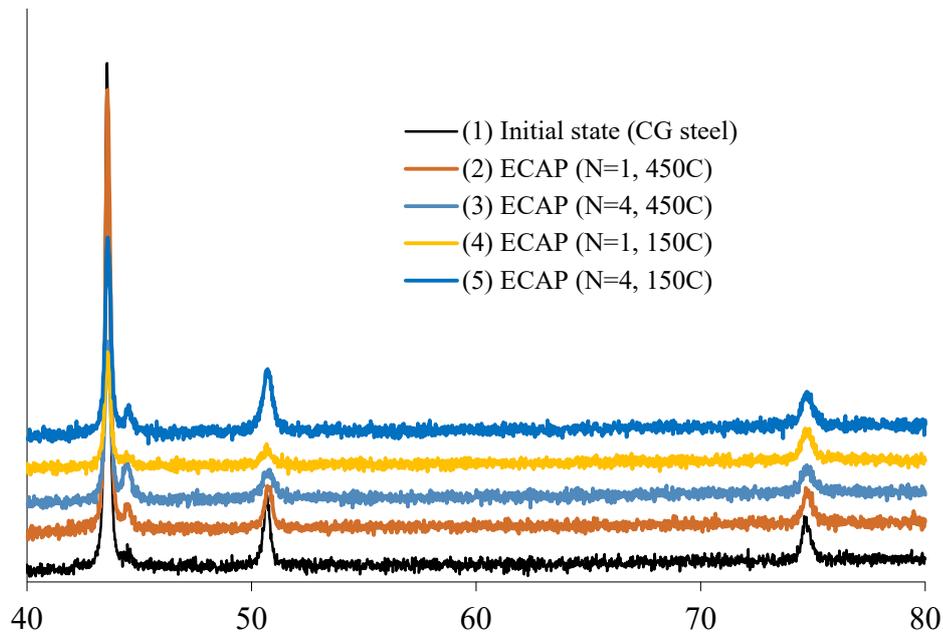

Figure 4. Results of XRD phase analysis of the steel samples in the initial state and after ECAP at 150 ºC and 450 ºC [44]

The heating of the samples up to the temperatures over 800 ºC led to the beginning of grain boundary migration and to nucleation of the σ-phase particles. Fig. 5 presents the results of *in-situ* investigation of evolution of microstructure of the UFG steel 321L (N = 4, 450 ºC). The samples of the UFG steel were heated from 300 ºC up to 800 ºC with a step of 100 ºC inside the TEM column; the holding time at every heating step was 1 hour. As one can see in Fig. 5, the UFG microstructure remained stable up to 700 ºC. High level of thermal stability of the UFG microstructure is provided by the nucleation of the σ-phase nanoparticles of ~10-20 nm in sizes (Fig. 5e, f, Fig. 6). The σ-phase nanoparticles are light-colored and have an equiaxial shapes (Fig. 6). The results of EDS analysis evidenced the nucleated nanoparticles to consist of Fe-Cr intermetallic compound (Fig. 6). When increasing the heating temperature up to 800 ºC (the holding time 1 hour), the sizes of the σ-phase particles increased up to 30-50 nm. As one can see in Fig. 5e, f and Fig. 6, the σ-phase nanoparticles were distributed uniformly over the section. Note, that the nucleation of the σ-phase particles during annealing of UFG austenitic steel 321 was reported earlier [46, 47].

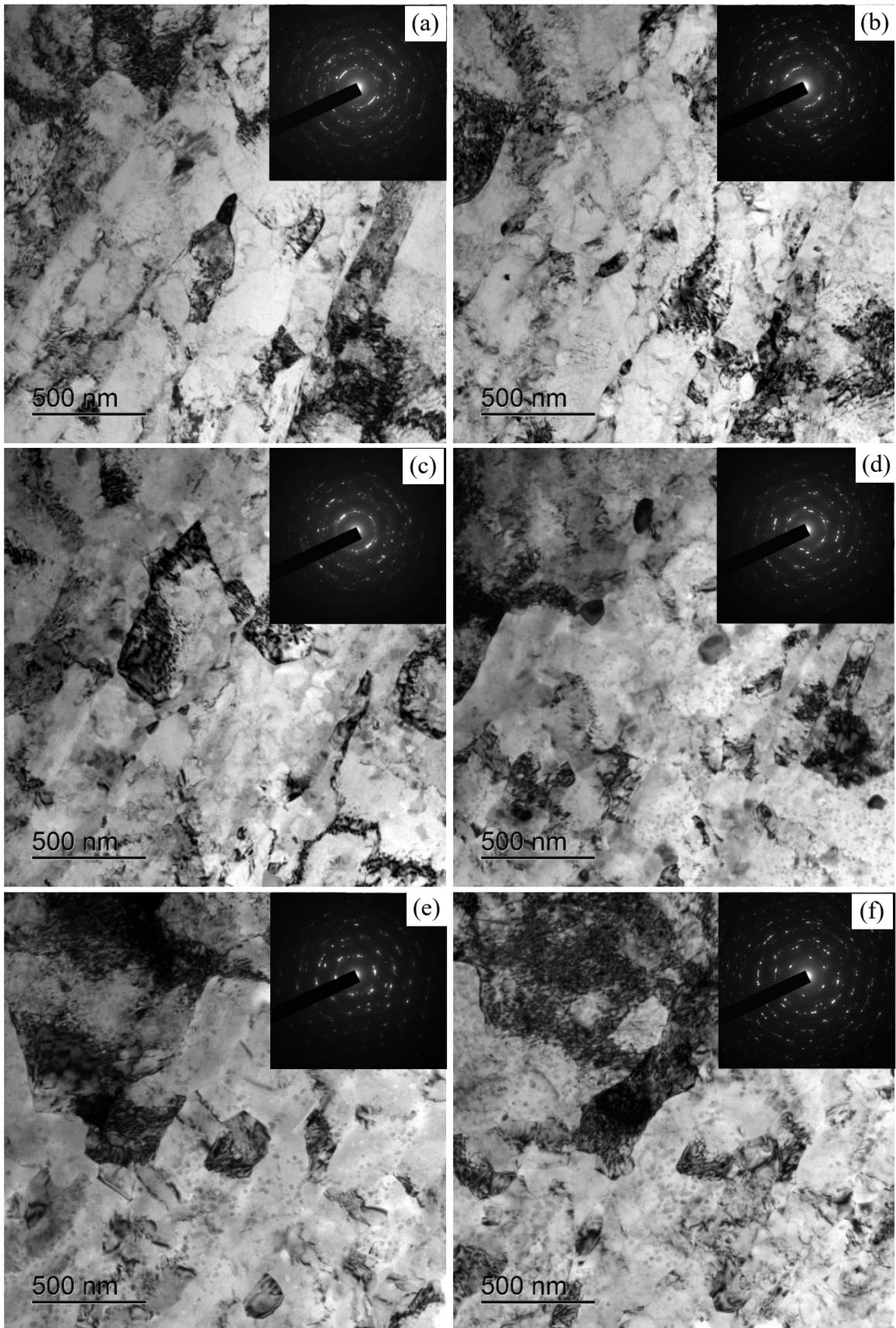

Figure 5. Results of *in-situ* investigations of the grain growth in UFG steel: a – heating up to 300 °C, holding 60 min; b – 500 °C, 60 min; c – 600 °C, 60 min; d – 700 °C, 60 min; e – 800 °C, 0 min; f – 800 °C, 60 min. TEM [44]

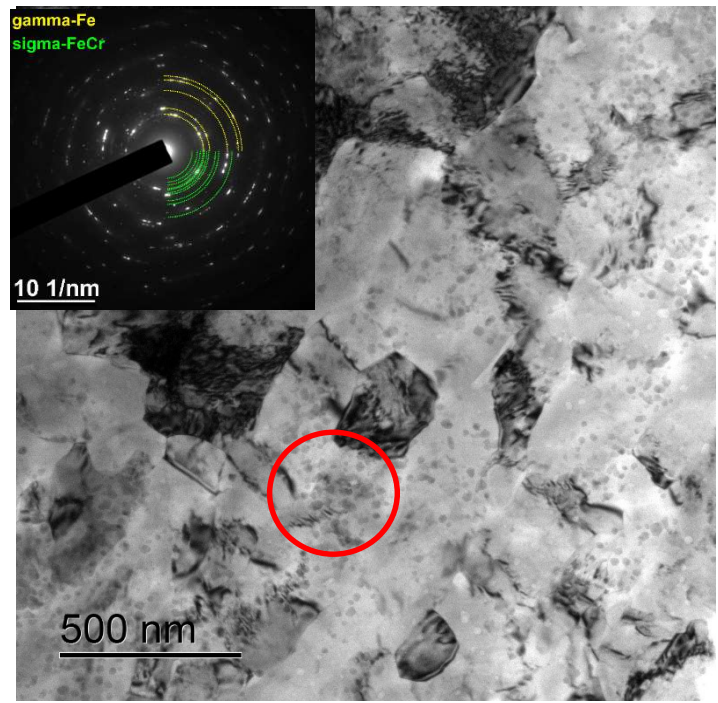

Figure 6. Nucleation of the σ-phase particles at the heating up of the UFG steel samples (N = 4, 450 °C) after at heating up to 700 °C and holding for 60 min. TEM [44]

3.2 Mechanical testing at RT

The stress–strain tension curves $\sigma(\varepsilon)$ of the CG austenitic steel specimens and of the UFG ones at RT are presented in Fig. 7. The $\sigma(\varepsilon)$ curve of the CG steel 321 had a conventional form, with a long strain hardening stage. The values of the ultimate strength of the investigated CG steel samples ($\sigma_B$ = 720 MPa) exceeded the usual ones for steel 321 ($\sigma_B$ = 490-520 MPa) essentially. In our opinion, high strength of steel 321 originates from the presence of the δ-ferrite particles (Fig. 2a, b), which played the role of stoppers of the deformation propagation inside the sample during the tensile testing.

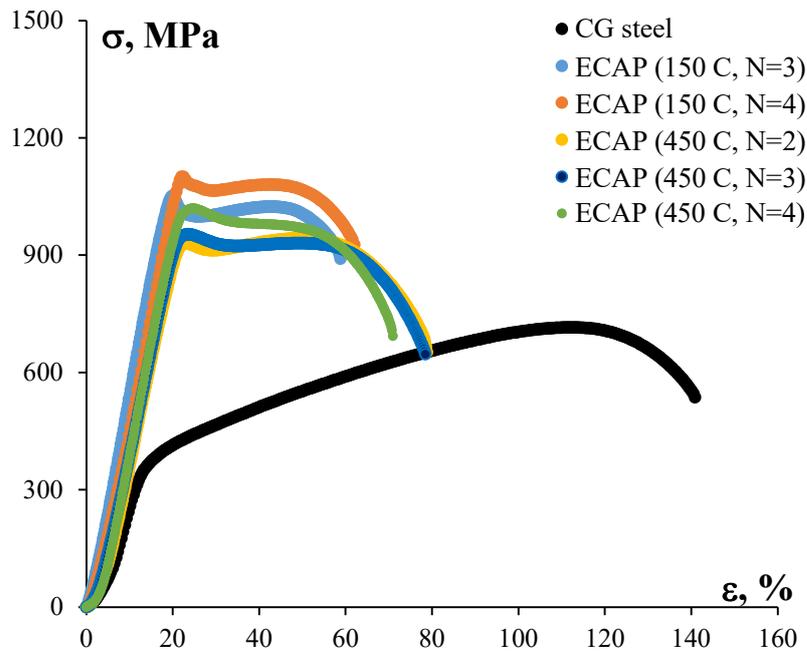

Figure 7. Stress–strain curves for the CG and UFG steel samples at RT

The $\sigma(\varepsilon)$ curves for the UFG steel samples at room temperature demonstrated the unstable plastic flow stage (Fig. 7). In the stress–strain curves $\sigma(\varepsilon)$ of the UFG steel specimens, there are a short stage of stable strain flow, which transforms into a stage of localization of the strain (Fig. 7). Increasing the ECAP temperature from 150 up to 450 °C resulted in an insufficient increase in the duration of the uniform strain stage. The duration of the stable strain flow stage for the samples made by ECAP at 450 °C was somewhat longer than for the ones made at 150 °C (Fig. 7). For all UFG samples a non-monotonous character of the curves $\sigma(\varepsilon)$ was observed at the initial stages – the stage of rapid stress decreasing in 30-40 MPa is changed by a long strain hardening stage or the strain stable flow of the UFG material (Fig. 7).

The tension tests of the CG and UFG steel specimens demonstrated the formation of the UFG structure by ECAP (N = 4, 150 °C) to result in a decreasing of the elongation to failure ($\delta$) of the steel 321 from 125 down to 45% and in an increasing of the ultimate strength from 720 up to 1100 MPa (Table 1). ECAP at higher temperature (450 °C) resulted in an insufficient decreasing of the ultimate strength down to 1020 MPa and in an increasing of the elongation up to $\delta \sim 60\%$.

The fractographic analysis revealed three characteristic zones on the fractures of the CG steel specimens and of the UFG ones after the tension tests (Fig. 8). These are: the fibrous zone, the radial one, and the break (cut) zone. It is worth noting that the cut zones in the CG steel occupied ~50 % of the whole fracture area. In the UFG steel after ECAP (N = 4), the cut zones occupied ~70%. So far, the formation of the UFG microstructure resulted in an increasing of the cut zone fraction of the fracture area and, hence, in a decreasing of the viscous component of the fracture.

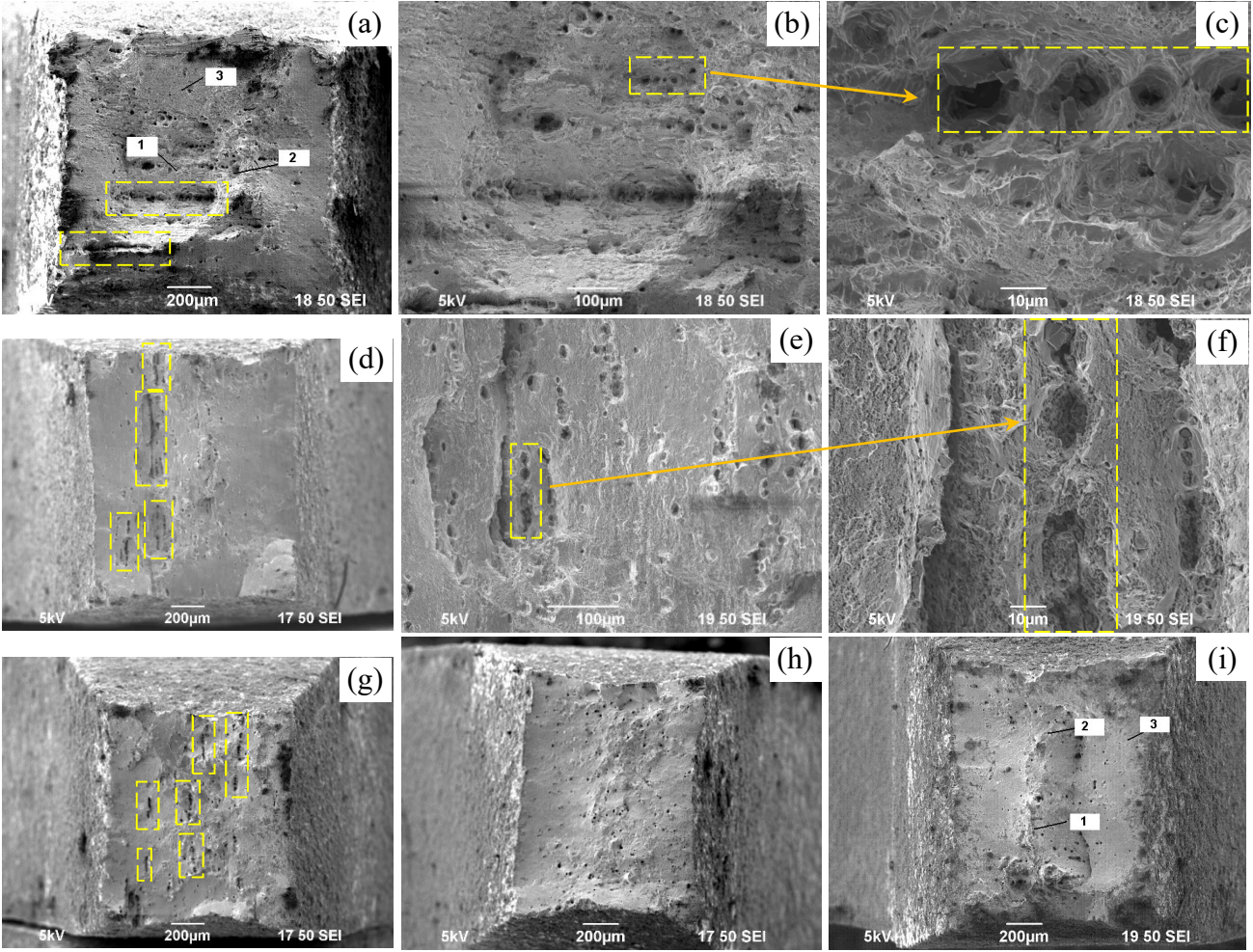

Figure 8. Fractographic analysis of the fractures of the CG (a, b, c) and UFG (d, e, f, g, h, i) steel samples after the tension tests at room temperature. Fig. 8 d, e, f present the fractures of the sample made by ECAP at 150 ºC (N = 4), Fig. 8g, h, i – the fractures of the sample made by ECAP at 450 ºC: (g) – N = 4, (h) – N = 3, (i) – N = 4. Zone 1 – the fibrous fracture zone; Zone 2 – the radial zone; Zone 3 – the cut zone. SEM

The elongated voids up to 0.5 mm long on the fractures of the CG samples are marked by yellow dashed lines in Fig. 8a. These voids formed probably in the areas of δ-ferrite location. Also, some separate voids of 10-20 μm in sizes are seen clearly on the fractures of the samples. These voids are formed likely due to the nonmetallic inclusions (Fig. 2). Some separate voids of 5-10 μm in sizes are seen mainly on the surfaces of the fractures of the UFG steel samples after ECAP at 450 °C (N = 3, 4) (Fig. 8g, h, i). Sometimes, the areas, which the fragmented δ-ferrite particles are present in are observed after ECAP at 150 °C. These areas are marked by white dashed lines in Fig. 8d, e, f. Probably, the ECAP temperature 150 °C is not enough for complete and efficient fragmentation of the δ-ferrite particles, which were present in the microstructure of the CG steel *ab initio* (Fig. 2a).

### 3.3 Tension testing at elevated temperatures

Table 1 presents the dependencies of the ultimate strength and of the elongation to failure on the testing temperature for the CG steel specimens and for the UFG ones obtained in different ECAP temperatures. Fig. 9 presents the curves $\sigma(\varepsilon)$ for the tension tests at elevated temperatures.

Table 1. Results of mechanical testing. The magnitudes of the ultimate strength ($\sigma_B$, MPa) and of the elongation to failure ($\delta$, %) for the samples of steel for the tension tests at $3.3 \cdot 10^{-3}$ s$^{-1}$

| $T_{test}$, °C | CG steel | | N = 2 | | | | N = 3 | | | | N = 4 | | | |
|---|---|---|---|---|---|---|---|---|---|---|---|---|---|---|
| | | | Temperature of the ECAP ($T_{ECAP}$) | | | | | | | | | | | |
| | | | 450 °C | | 150 °C | | 450 °C | | 150 °C | | 450 °C | | | |
| | $\sigma_B$ | $\delta$ | $\sigma_B$ | $\delta$ | $\sigma_B$ | $\delta$ | $\sigma_B$ | $\delta$ | $\sigma_B$ | $\delta$ | $\sigma_B$ | $\delta$ | | |
| RT | 720 | 125 | 950 | 70 | 1100 | 40 | 950 | 65 | 1100 | 45 | 1020 | 60 | | |
| 450 | 420 | 65 | 800 | 40 | 870 | 35 | 720 | 20 | 920 | 22 | 760 | 30 | | |
| 600 | 350 | 65 | 650 | 50 | 600 | 50 | 600 | 48 | 630 | 45 | 640 | 45 | | |
| 750 | 250 | 70 | - | - | 120 | 250 | 290 | 105 | 240 | 185 | 290 | 120 | | |
| 800 | 220 | 75 | 250 | 110 | 150 | 200 | 200 | 150 | 152 | 220 | 205 | 160 | | |
| 900 | - | - | - | - | - | - | 98 | 190 | - | - | - | - | | |

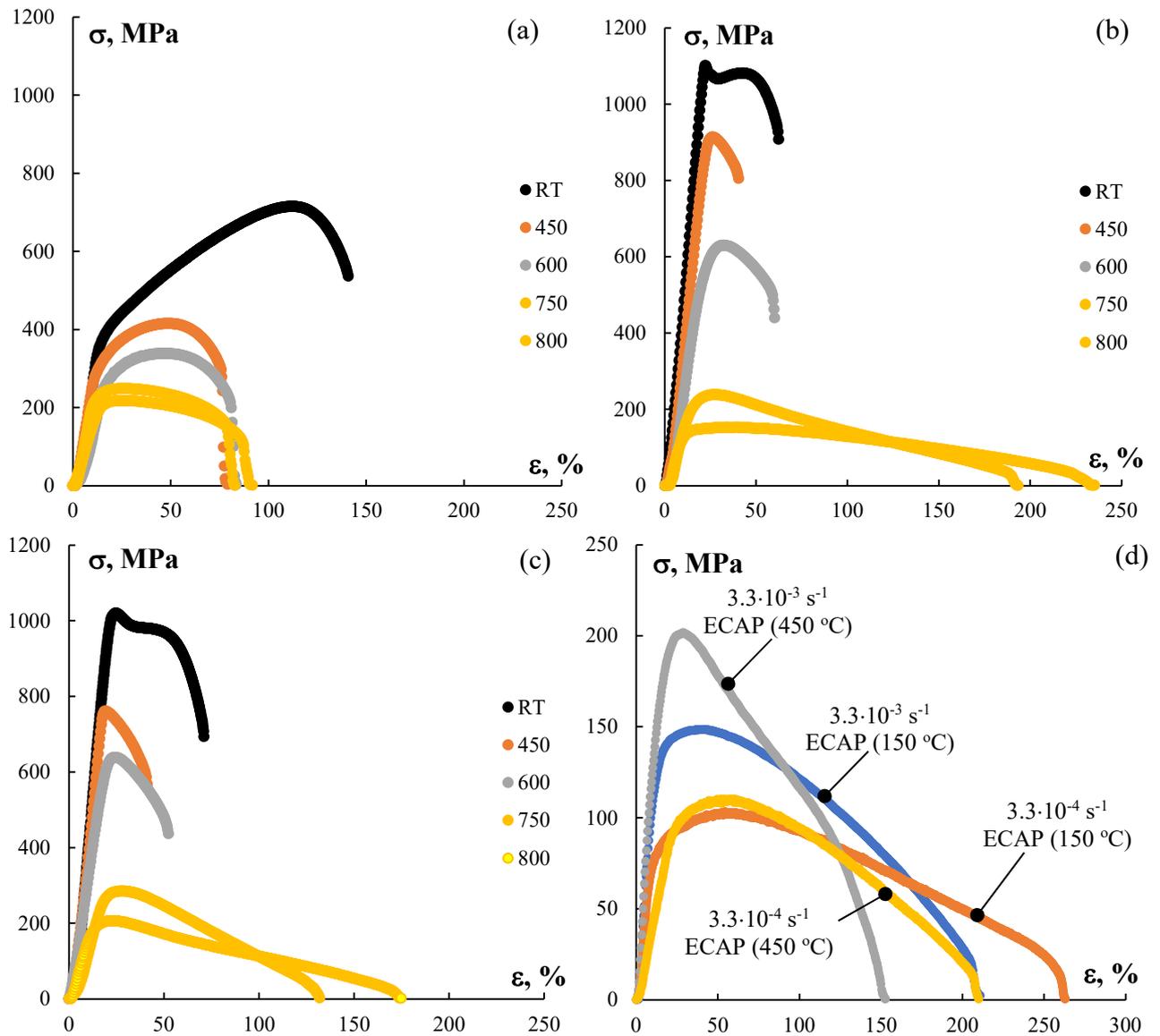

Figure 9. Stress–strain curves for the CG (a) and UFG (b, c, d) steel at elevated temperatures: (a) CG steel; (b) UFG steel (N = 4, 150 ºC), (c) UFG steel (N = 4, 450 ºC), (d) analysis of the effect of strain rate on the elongation to failure of the UFG steel samples (N = 3) at 800 ºC

The stress–strain curves $\sigma(\varepsilon)$ for the CG steel specimens had the form typical for high-ductility materials (Fig. 9a). The duration of the localized strain stage was much smaller than of the uniform strain one. The curves $\sigma(\varepsilon)$ for the UFG steel specimens at the testing temperatures of 750 and 800 ºC had the form typical for highly ductility materials – the stage of an insufficient strain hardening transformed into a long state of stable strain flow (Figs. 9b, c). The analysis of the curves presented in Fig. 9 demonstrated the increasing of the temperature from RT up to 750 °C to result in a monotonous decreasing of the ultimate strength (the flow stress) from 720 MPa down to 250 MPa for the CG steel and from 950–1100 MPa down to 240–290 MPa for the UFG steel, respectively. No unstable plastic flow regions were observed in the high-temperature $\sigma(\varepsilon)$ curves for the CG and UFG steels (Fig. 9).

Note that the increasing of the testing temperature resulted in a nonmonotonous variation of the elongation to failure for the UFG steel that differs from the same dependencies for the CG steel. The analysis of the data presented in Table 1 shows the elongation to failure for the CG steel to decrease monotonously from 125% down to 70% with increasing testing temperature from RT up to 750 °C. At the test temperature 800 °C, the plasticity of the CG steel increased insufficiently ($\delta \sim$ 90%). As it has been already mentioned in Introduction, the non-monotonous (with a minimum) character of the dependence $\delta(T)$ is typical for the metastable CG austenitic steel.

The fractures of the CG steel samples after the tension tests at elevated had viscous character (Fig. 10). The central fibrous zones of the sample fractures (Zone 1) after tests at 450-600 °C comprised sets of very small pits of 5-10 μm in sizes (Fig. 10a-10d). There were few deep pores in the fibrous zones, the shapes of which were close to the spherical ones (Fig. 10b, d). After the tests at 750-800 °C, the areas of the radial zones of the fractures (Zone 2) increased essentially whereas the areas of the cut zones (Zone 3) tend to zero. The sizes and depths of the pits in the fibrous zones of the CG steel fractures increased essentially after the tests at 750-800 °C, the coalescence of the pits into the elongated voids up to 50 μm long was observed. It is important to note that the fractures of the CG steel samples remained viscous even after the tension tests at 750-800 °C, although the ductility of the CG steel at these temperatures was lower than the one at room temperature.

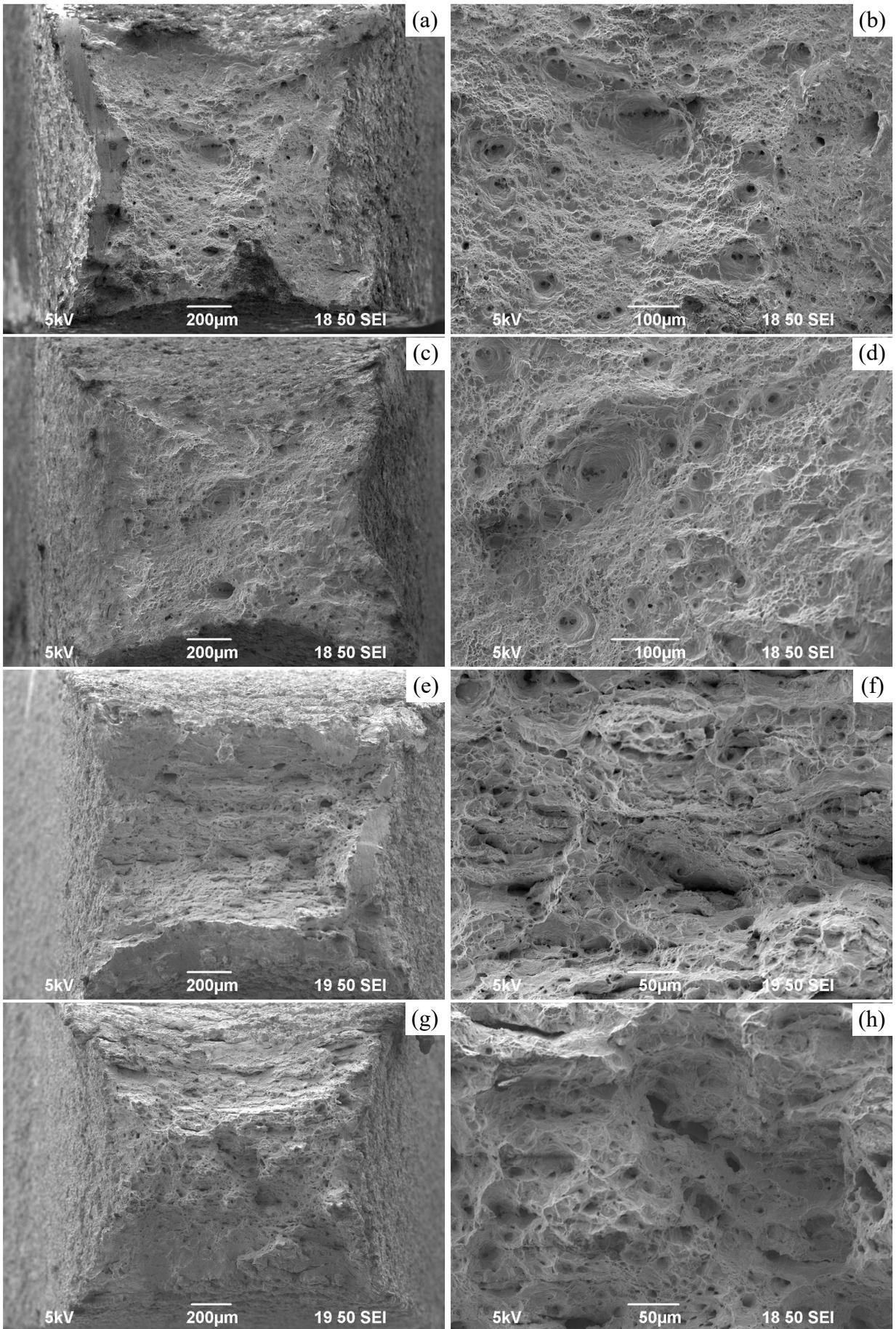

Figure 10. Fractographic analysis of the CG steel fractures after the superplasticity tests at 450 °C (a, b), 600 °C (c, d), 750 °C (e, f), 800 °C (g, h). SEM

The character of the dependence δ(T) for the UFG steel was more complex – the elongation to failure decreased insufficiently with increasing testing temperature from RT up to 450 °C. The increasing of the number of ECAP cycle resulted in a decrease in the elongation to failure at 450 °C. At further increasing of the temperature from 450 °C up to 750–800 °C, the elongation to failure of the UFG steel increased and was several times higher than the δ of the CG steel. For the UFG steel specimens obtained by ECAP at 150 °C, the elongation to failure at the testing temperature of 750 °C reached 250%. Further increasing of the testing temperature resulted in a decreasing of the elongating to failure for the UFG steel specimens again.

To determine the magnitude of the strain rate sensitivity coefficient $m$, the samples made by ECAP (N = 3, 4) at 150 °C and 450 °C were investigated. For example, Fig. 9d presents the curves σ(ε) for the UFG steel 321 samples (ECAP, N = 3) deformed at 800 °C and different strain rates $\dot{\varepsilon}$. The analysis of the σ(ε) curves has shown the decreasing of the strain rate from $3.3 \cdot 10^{-3}$ s$^{-1}$ down to $3.3 \cdot 10^{-4}$ s$^{-1}$ to result in a decrease in the flow stress and in an increase of the elongation to failure. No unstable Lüders deformation ranges were observed in the σ(ε) curves at 800 °C in the rate range investigated. The magnitude of the strain rate sensitivity coefficient $m$ was determined from the slope of the $\sigma_b(\dot{\varepsilon})$ dependence plotted in the logarithmic axes. For the UFG steel samples made by ECAP at 150 and 450 °C, the magnitudes of the coefficient $m$ were 0.16-0.18 and 0.26-0.29, respectively.

The metallographic investigations of the UFG steel samples after testing at elevated temperatures evidenced an intensive pore formation in the destruction areas (Fig. 1b). Note that large elongated spindle shaped pores are present in the UFG steel samples made by ECAP at elevated temperatures (Fig. 1b). The lengths of such pores reach ~100-150 μm (Fig. 1b). In the fracture zone, small micron-sized pores distributed uniformly over the UFG sample surfaces are present also.

The fractographic analysis of the fractures (Figs. 11, 12) demonstrated the areas of the fibrous zones and of the radial ones to increase and the areas of the cut zones – to decrease with increasing testing temperature. At the testing temperature of 600 °C, the cur zone area didn't exceed 5–10% of the whole fracture area. For the UFG steels ($T_{ECAP}$ = 450 °C), the cut zones were absent that also evidences an increased ductility of the UFG material as compared to the coarse-grained state.

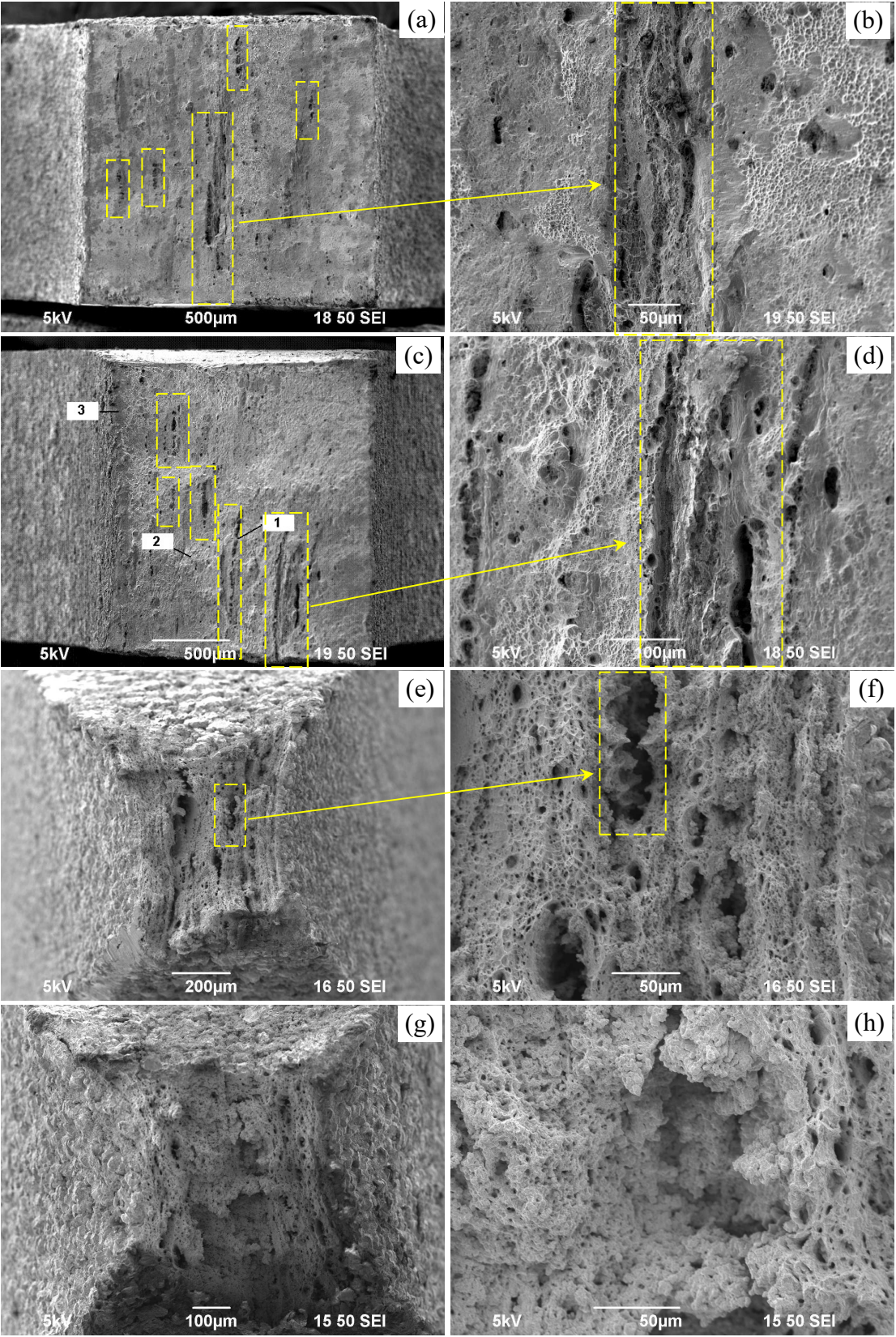

Figure 11. Fractographic analysis of the UFG steel (ECAP, N = 4 at 150 °C) fractures after the superplasticity tests at 450 °C (a, b), 600 °C (c, d), 750 °C (e, f), and 800 °C (g, h). SEM

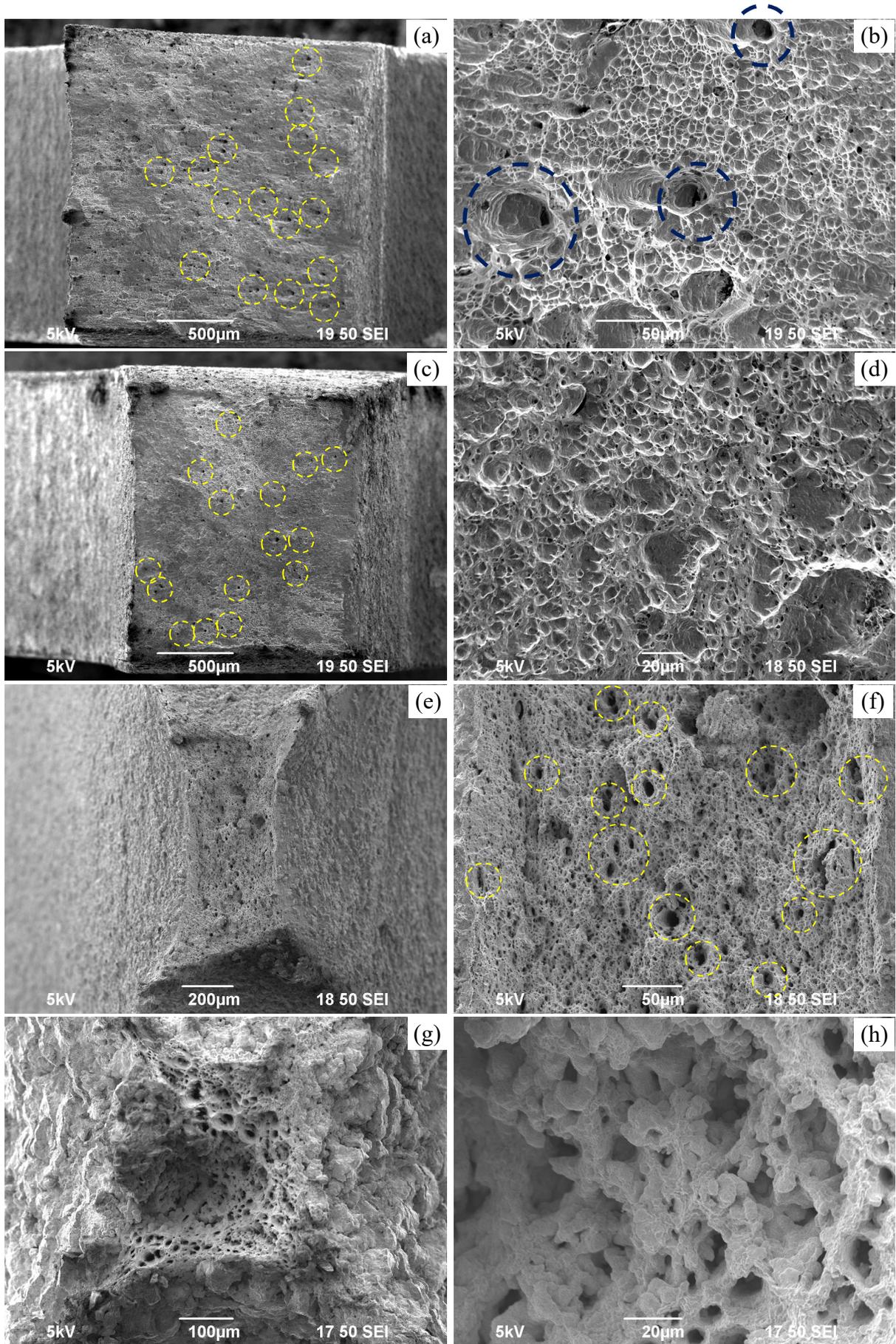

Figure 12. Fractographic analysis of the UFG steel (ECAP, N = 3 at 450 °C) fractures after the superplasticity tests at 450 °C (a, b), 600 °C (c, d), 850 °C (e, f), and 900 °C (g, h). SEM

It is worth noting that the character of the fractures of the UFG steel samples made by ECAP at 150 °C (Fig. 11) and 450 °C (Fig. 12) is different. After the tension tests at low temperatures (450-600 °C), the elongated voids up to 500 μm long remained on the fracture surfaces of the UFG steel samples (ECAP at 150 °C) (Fig. 11a-11d). The largest voids are marked by the dashed lines in Fig. 11 for clarity. Similar voids being likely the destruction regions of δ-ferrite were observed on the fractures of the CG and UFG steel after the tests at room temperature (Fig. 8). The sampled cross section areas decreased with increasing test temperature (Fig. 11e, g); the elongated voids transformed into the equiaxial pores (Fig. 11g, h).

Fig. 12 presents the results of the SEM investigations of the fractures of the UFG steel samples obtained by ECAP at 450 °C. There were few large pores and shallow pits in the central fibrous zones of the fracture surfaces of the UFG steel samples (ECAP at 450 °C) after the superplasticity tests at 450-600 °C. Large pores in Fig. 12a-12d are outlined by dashed lies. The large elongated voids were observed in the samples, the fine-grained microstructure in which was formed at N = 1 and N = 2 ECAP cycles at 450 °C, but were absent in the samples after N = 3 and N = 4 ECAP cycles at 450 °C. The areas of the radial zones of the fractures decreased with increasing testing temperature up to 800 °C whereas the fibrous zones of the fractures consisted of pits and pores of various sizes completely (Fig. 12e, f). The sizes of some pores reach 20 μm. The largest pores are outlined by dashed lines in Fig. 12f. At further increasing of the tension temperature up to 900 °C, the cross section areas of the samples decreased essentially (Fig. 12g); the fractures had fully viscous character and comprised an ensemble of deep pores.

In the non-deformed part of the CG steel 321 samples, a uniform austenite microstructure remained after tensile testing at elevated temperatures (Fig. 13a). The mean austenite grain sizes almost didn't depend on the testing temperature. Pores of submicron and micron sizes are formed at the austenite grain boundaries (Fig. 13b). Large twin are seen inside the non-deformed austenite grains. In the deformed parts of the CG samples, the elongated reversed austenite grains were observed inside some equiaxial austenite grains (Fig. 13c, d). These elongated austenite grains were formed probably as a result of transformation of the α′-martensite particles.

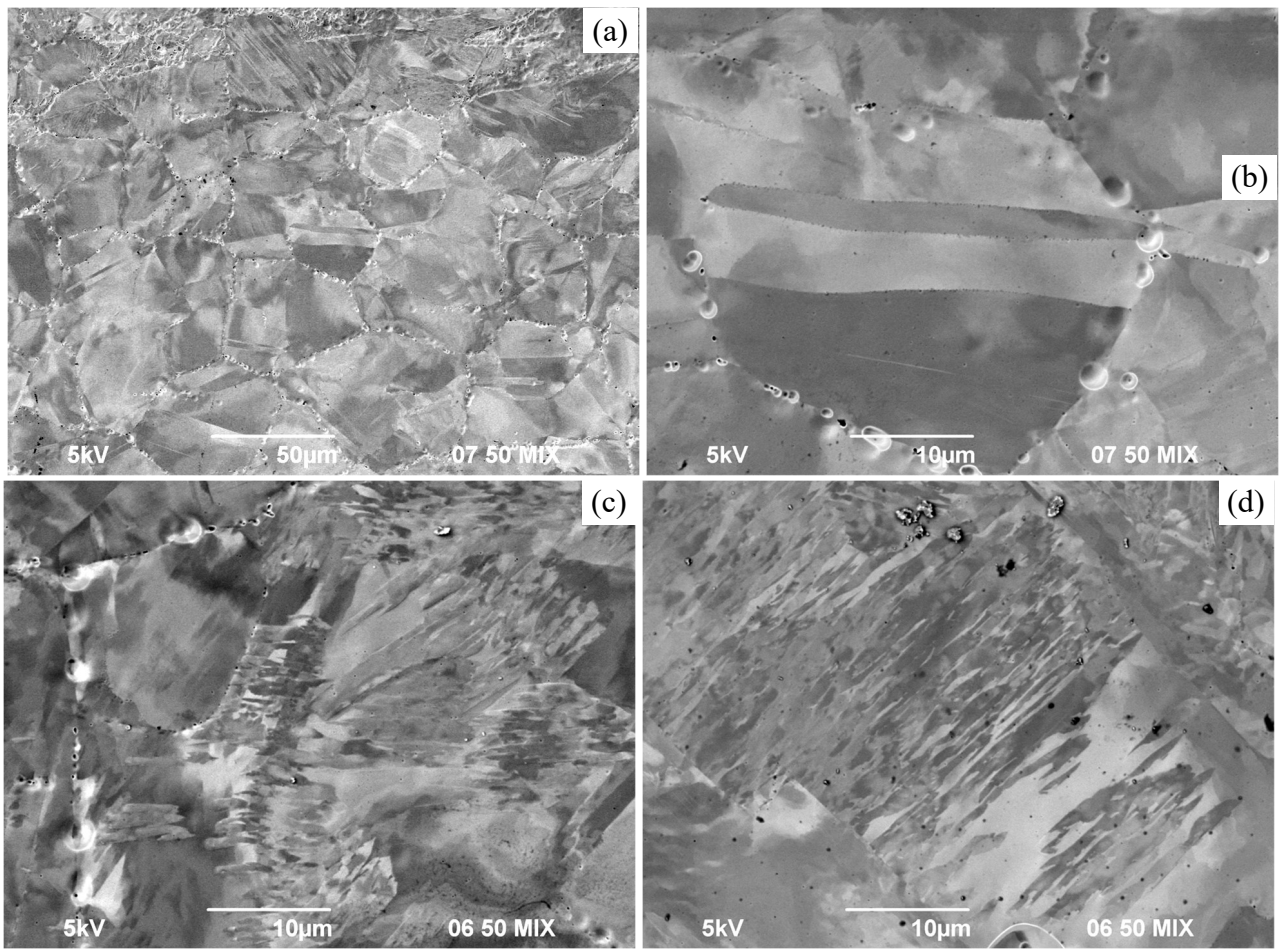

Figure 13. Results of the microstructure investigations of the nondeformed areas (a, b) and of the deformed ones (c, d) of the specimens CG steel after the tension tests at 750 ºC and 800 ºC (a, b, d). SEM

The laws of evolution of the microstructure of the UFG steel during the deformation were more complex. At reduced testing temperatures (700-750 ºC) close to the temperature of recrystallization, the main grain sizes in the deformed zones were slightly greater then the ones in the non-deformed zones (Fig. 14a, b). At elevated temperatures (800-900 ºC), the mean grain sizes in the deformed zones were slightly smaller or comparable with the grain sizes in the non-deformed zones (Fig. 14c, d, Table 2). The grains preserved the equiaxial shapes. No elongated austenite grains expressed clearly were found when etching the deformed zones of the samples (Fig. 14b, d). This result allows suggesting the strain-induced grain growth to take place at reduced deformation temperatures whereas at higher deformation temperatures the dynamic recrystallization starts.

Table 2. Microhardness of the samples of steel after tension testing at different temperatures at $3.3 \cdot 10^{-3}$ s$^{-1}$. The mean recrystallized grain sizes [d, μm] for the samples tested at 800 and 900 °C are given in braces

| $T_{test}$, °C | CG steel | | N = 2 | | | | N = 3 | | | | N = 4 | |
|---|---|---|---|---|---|---|---|---|---|---|---|---|
| | | | \multicolumn{10}{c|}{Temperature of the ECAP ($T_{ECAP}$)} |
| | | | 450 °C | | 150 °C | | 450 °C | | 150 °C | | 450 °C | |
| | Zone I | Zone II | Zone I | Zone II | Zone I | Zone II | Zone I | Zone II | Zone I | Zone II | Zone I | Zone II |
| RT | 2.15 | 3.71 | 3.49 | 4.34 | 3.91 | 4.45 | 3.47 | 4.38 | 3.99 | 4.50 | 3.51 | 4.46 |
| 450 | 1.87 | 2.95 | - | - | - | - | 3.64 | 3.55 | 4.07 | 4.13 | 3.76 | 3.75 |
| 600 | 1.88 | 2.68 | 2.94 | 3.38 | 4.05 | 3.63 | 3.56 | 3.49 | 4.25 | 3.75 | 3.77 | 3.60 |
| 750 | 1.76 | 2.38 | - | - | - | - | 2.79 (~0.5) | 2.77 (~1) | 3.33 | 2.52 | 3.38 | 2.71 |
| 800 | 1.84 (33) | 2.26 (41) | 2.40 (21.5) | 2.54 (2.0) | 2.13 (2.7) | 2.39 (2.3) | 2.15 (2.9) | 2.48 (1.6) | 2.96 (2.5) | 3.05 (1.9) | 2.21 (2.6) | 2.29 (1.6) |
| 900 | - | - | - | - | - | - | 1.90 (6.8) | 1.99 (4.4) | - | - | - | - |

Note: Zone I is the non-deformed one, Zone II is the deformed one (the fracture zone) (Fig. 1)

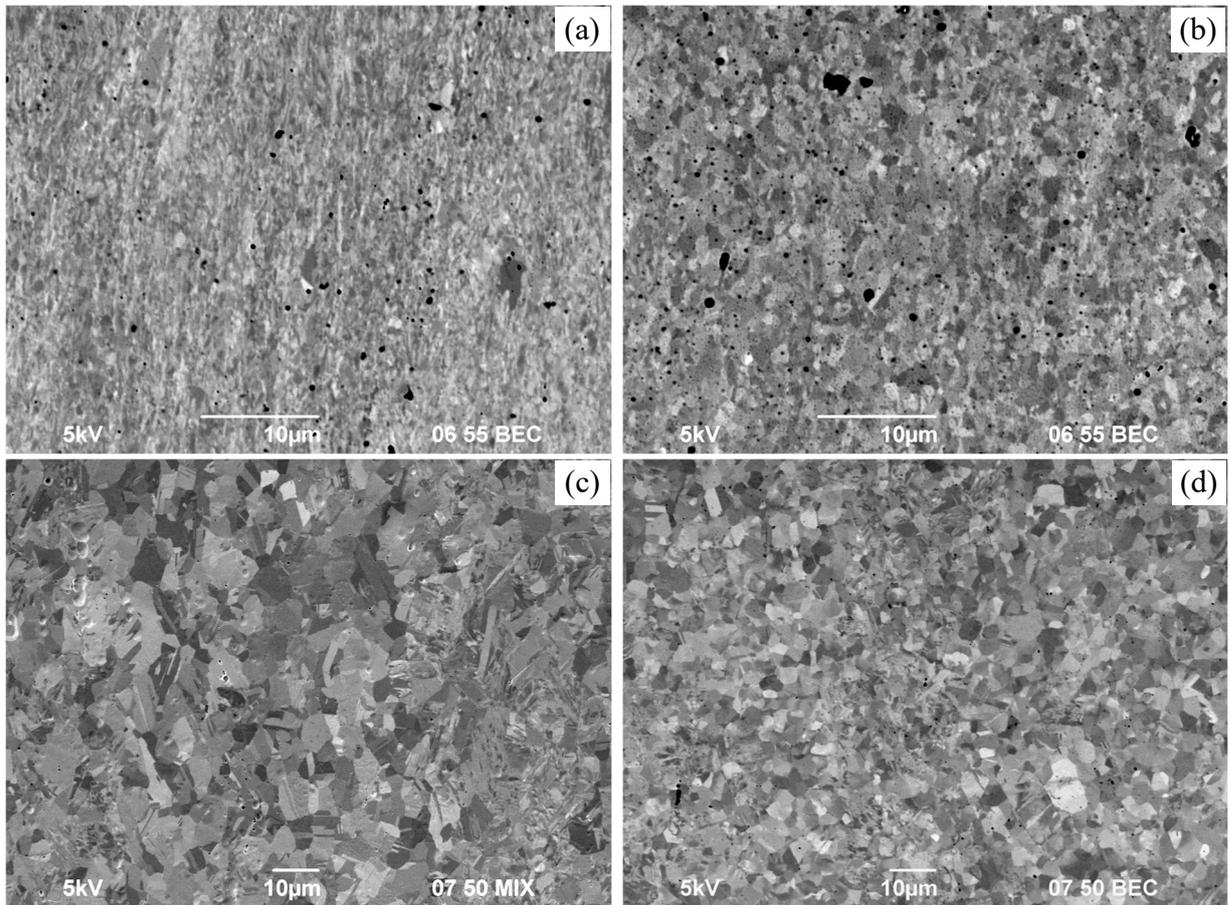

Figure 14. Results of the microstructure investigations of the nondeformed areas (a, c) and of the deformed ones (b, d) of the specimens UFG steel (ECAP, N = 3, 450 °C) after the tension tests at 750 °C (a, b) and 900 °C (c, d). SEM

The microhardness testing results (Table 2) evidence the recrystallization at the high-temperature testing of the UFG steel. The microhardness measurements of the specimens after the tension tests demonstrated the increasing of the testing temperature from 450 up to 800 °C to result in a decreasing of the microhardness both in the deformed areas and in the non-deformed ones. This conclusion is supported by the results of the microstructure investigations in the deformed areas of the UFG steel specimens and in the non-deformed ones after the tension tests (Fig. 14). As one can see from Fig. 12 and from the data presented in Table 2, the testing at 800 °C resulted in the formation of a well uniform fine-grained microstructure. No essential grain growth was observed. The mean grain sizes in the deformed parts were slightly smaller than in the non-deformed ones.

**Discussion**

First, let us discuss the question of the origin of the cavitation fracture of the UFG steels in the superplasticity conditions. The cavitation limits the possibility to achieve higher superplastic deformation characteristics of UFG steels 321. Cavitation is one of known mechanisms of destruction of materials in the superplasticity conditions [48, 49].

A model of pore formation at the second phase particles located in the triple grain junctions in the superplasticity conditions was proposed in [49, 50]. According to [49, 50], the pore nucleation in the triple grain boundary junctions is caused by the accumulation of the normal components of the delocalized dislocations. The related elastic energy and internal stresses of the defect layer formed at the particle – matrix interphase boundary increase in the course of superplastic deformation [49]. A detailed description of these defects is given in [51, 52]. A defect nucleated at a particle with the radius R located inside a grain boundary, in the first approximation, can be described as a disclination loop with the radius R and the power $\omega(t)$. The power of the disclination loop $\omega(\tau)$ grows proportionally to the number of defects falling onto the grain boundary during the superplastic deformation: $\omega(t) = \psi_1 \dot{\varepsilon}_v t$ [49, 51] where $\psi_1$ is a geometrical factor, $\dot{\varepsilon}_v$ is the intergranular strain rate, and t is the strain time. The power $\omega(t)$ grows with accumulation of the defects at the grain boundaries and, correspondingly, related elastic energy of the disclination loop increases. As it has been shown in [49], at certain critical power $\omega^*$, the related elastic energy of the loop becomes so high that it becomes favorable energetically to the grain boundary to release from the source of this excess energy. At reduced deformation temperatures, it would lead to formation of microcracks near the second phase particles. At elevated deformation temperatures, the relaxation of the accumulated can go via the formation of micropores at the particle – metal interphase boundaries [49].

As it has been shown above, steel 321 in the initial state contains large non-metallic particles TiN and Ti(C,N) (Fig. 2). Also, the nucleation of the σ-phase particles was observed when heating the UFG steel. Cutting such large particles by the lattice dislocations is difficult, and the disclination-type defects would form around these ones in the course superplastic deformation. In our opinion, large TiN and Ti(C,N) particles (Fig. 2a, b) can be the sources of formation of the large pores in the superplastic deformation conditions while the σ-phase particles can be the origin of the micropore formation (see [53]). Usually, the σ-phase particles in the austenitic steel can form at the transformation (decomposition) of δ-ferrite [1, 54], but the formation rate of the σ-phase particles is very low in normal conditions [1, 55, 56]. Some works evidence a possibility of forming of the σ-phase particles at the decomposition of the α′-phase [57] or directly from the γ-phase [58, 59]. In the case of formation of the σ-phase at the austenite grain boundaries, its formation rate is limited by the intensity of chromium diffusion in the grain boundaries [56, 58]. Dislocations also can be the regions of nucleation and growth of the σ-phase particles [60]. Simultaneous transformation of the δ-ferrite particles and the nonequilibrium state of the austenite crystal lattice, which a large number of defects are present in are the most probable origins of accelerated nucleation of the σ-phase particles in the UFG steels. Possible effect of SPD, which leads to the acceleration of the diffusion processes in the materials should be noted also [30, 61].

The negative effect of the σ-phase on the high-temperature plasticity of the austenite steels was also noted in [62-64]. The presence of the σ-phase particles can lead to the pore formation and cavitation as in the case of ferrite particles in austenite [65]. Note also that the σ-phase can form at the hot deformation (see, for example, [1, 63, 66, 67]), in particular – in superplasticity [68, 69]. At that, the increasing of the strain magnitude and forming a high density of various defects in austenite promotes the increase in the σ-phase content [70-73]. The growth of the pores in the course of superplastic deformation would go proportionally to the strain degree and strain rate [49, 50]. Upon achieving some critical sizes, such pores become the sources of the fracture of the UFG steel samples.

Now, let us analyze the superplastic deformation mechanisms in the UFG steel 321.

The main rheological equation of the superplastic flow is usually presented in the form [51, 73]:

$$\dot{\varepsilon} = A\left(\sigma^*/G\right)^{1/m}(b/d)^p(D_{sp}/b^2)(G\Omega/kT), \qquad (1)$$

where *m* is the strain rate sensitivity coefficient for the flow stress, the magnitude of which depends on the strain rate $\dot{\varepsilon}$, *p* is a numerical parameter equal to 2 or 3; $D_{SP} = D_0 \exp(Q_{SP}/kT)$ is the diffusion coefficient in the superplastic conditions, $Q_{SP}$ is the activation energy of the superplastic flow, $\sigma^*$ is the effective superplastic flow stress (usually, it is supposed that $\sigma^* = \sigma_b$), G is the shear modulus, $\Omega$ is the atomic volume, k is the Boltzmann constant, and b is the Burgers vector.

In the case when the grain boundary deformation ($\dot{\varepsilon}_b$) mechanism and the internal ($\dot{\varepsilon}_v$) one are present simultaneously in the superplastic deformation conditions, the rheological equation is often presented in the form:

$$\dot{\varepsilon} = \dot{\varepsilon}_b + \dot{\varepsilon}_v. \qquad (2)$$

To describe the grain boundary strain rate $\dot{\varepsilon}_b$, the equation for grain boundary sliding [51, 73] is employed usually:

$$\dot{\varepsilon}_b = A_b(\sigma/G)^2 (b/d)^2 (G\Omega/kT)(\delta D_b/b^3), \qquad (3)$$

where $A_b$ is a numerical factor equal to ~100, $\delta = 2b$ is the grain boundary width, and $D_b$ is the grain boundary diffusion coefficient. The possibility of realization of the GBS mechanism in the fine-grained austenite steel at elevated straining temperature was demonstrated in [35, 74].

To calculate the intragranular strain rate $\dot{\varepsilon}_v$, the equation for the power-law creep [73, 75] is used often:

$$\dot{\varepsilon}_v = A_v(\sigma/G)^n (D_v/b^2)(G\Omega/kT), \qquad (4)$$

where $A_v$ is the Dorn constant, $D_v$ is the self diffusion coefficient in the crystal lattice.

During the deformation of the fine-grained alloys in non-optimal for superplasticity conditions, the intragranular deformation ($\dot{\varepsilon}_v$) gives the major contribution into total strain rate ($\dot{\varepsilon} = \dot{\varepsilon}_v + \dot{\varepsilon}_b$) where $\dot{\varepsilon}_b$ is the GBS rate [35, 74]. It will result in low values of the strain rate sensitivity coefficient (*m* < 0.5) [74]. Note that for the UFG steel 321L the value of the coefficient *m* doesn't reach the optimal value *m* = 0.5. This allows suggesting the contribution of the intragranular deformation ($\dot{\varepsilon}_v$) during the hot deformation of the UFG steel specimens to be large enough.

Using Equation (1), one can estimate the effective activation energy of the superplastic deformation for the UFG steel 321. According to (1), the magnitude of $Q_{SP}$ [in $kT_m$] can be determined from the slope of the dependence $\ln(\sigma_B)/m - T_m/T$ where $T_m = 1810$ K is the melting point of iron. As one can see in Fig. 15, the dependencies $\ln(\sigma_B)/m - T_m/T$ in the temperature range 600-800 °C can be fitted by straight lines with a good precision. The reliability coefficient of the linear fit for all lines $R^2 > 0.90$. The analysis shows the magnitude $Q_{SP}$ for the UFG steel samples obtained by ECAP at 450 °C ($m = 0.27$) is 9.5-11.7 $kT_m$ (~ 143-176 kJ/mol) and is close to the activation energy of the grain boundary diffusion in the austenite ($Q_b \sim 159$ kJ/mol [75]). For the UFG steel 321 samples obtained by ECAP at 150 °C ($m = 0.16$), the activation energy of superplastic deformation is ~22.1-26.7 $kT_m$ (~ 332-402 kJ/mol). The value $Q_{SP} \sim 332\text{-}402$ kJ/mol exceeds the activation energy of diffusion in the austenite crystal lattice ($Q_v \sim 270$ kJ/mol [75]) but is close to the activation energy for the case of hot deformation of metastable austenitic steel (316, 321, etc.) [75]. It is interesting to note that the mean grain sizes in the deformed zones of the samples (Zones II) made at $T_{ECAP} = 150$ °C is greater than the mean grain sizes in the deformed zones parts of the samples made at $T_{ECAP} = 450$ °C. According to (1), at $\dot{\varepsilon} = $ const, it should lead to an increase in the flow stress $\sigma_y$ (see also [52, 73]). One can see from Table 1 that at $T_{SP} = 750\text{-}800$ °C the flow stress $\sigma_B$ for the UFG steel samples made at $T_{ECAP} = 150$ °C is less than the one for the samples made at $T_{ECAP} = 450$ °C.

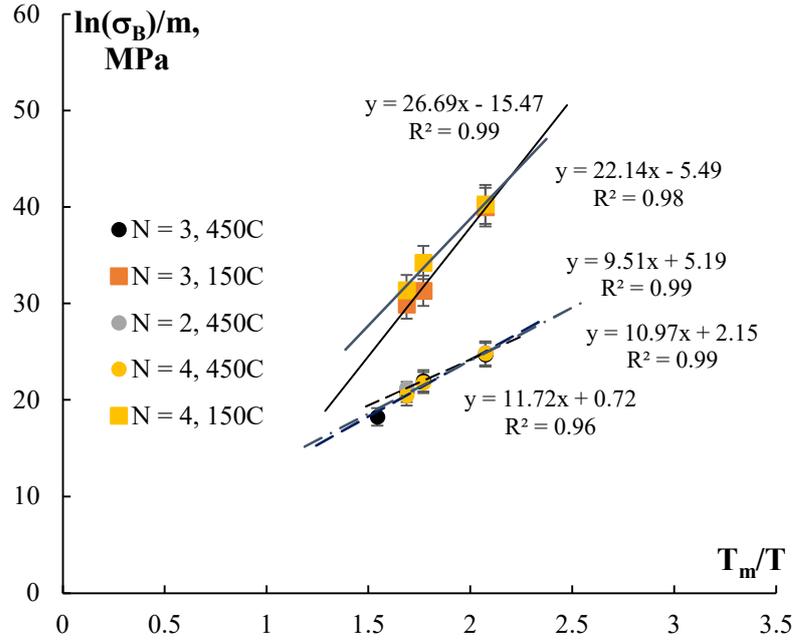

Figure 15. Temperature dependence of the flow stress plotted in the $\ln(\sigma_B)/m - T_m/T$ axes. Calculation of the activation energy of superplastic deformation in UFG steel

In our opinion, the origin of this contradiction is the effect of the σ-phase nanoparticles, which prevent the grain boundary migration. More intensive nucleation of the σ-phase particles during the superplastic deformation of the UFG steel 321 samples ($T_{ECAP} = 450$ °C) leads to the decrease in the grain growth rate according to Zener equation (see [76]) and, as a consequence, to the increase in the GBS rate according to Equation (1).

The intensive grain growth during superplastic deformation of the UFG steel 321 samples ($T_{ECAP} = 150$ °C) leads to the decreasing of GBS in accordance to Equation (3), and he contribution of the power law creep becomes essential (see Equation (2)). This leads to an increase in the superplasticity activation energy so that it become close to the activation energy of the power law creep (see Fig. 14). From the condition $\dot{\varepsilon}_b = \dot{\varepsilon}_v$, one can find the critical grain size $d^*$, which the dominant deformation mechanism in the austenitic steel changes at:

$$d^* = \sqrt{(A_b/A_v)b(G/\sigma)(\delta D_b/D_v)}. \tag{5}$$

At $A_b = 100$, $A_v = 4.3 \cdot 10^5$ [75], $b = 2.58 \cdot 10^{-10}$ m, $G = 81$ GPa, $\delta D_0 = 7.5 \cdot 10^{-14}$ m$^3$/s [75], $Q_b = 159$ kJ/mol [75] $\sim 10.6$ $kT_m$, $D_{v0} = 1.8 \cdot 10^{-5}$ m$^2$/s [75], $T = 1073$ K, one obtains the critical grain size $d^* \sim 0.16$ μm. This value is an order of magnitude smaller than the mean grain sizes observed experimentally (Table 2).

In our opinion, the disagreement between the calculation results and the experimental data originates mainly from the nonequilibrium state of the grain boundaries in the UFG alloys in the superplastic deformation conditions [30, 77, 78]. During the superplastic deformation, the grain boundaries in the UFG metals contain an increased density of defects: orientation mismatch dislocations (OMDs) and products of delocalization of these ones – the tangential components of the delocalized dislocations. The defects trapped by the grain boundaries enhance the free (excess) volume of the grain boundaries and lead to the decreasing of the activation energy of the grain boundary diffusion [30, 77, 78]. A good agreement between the results of calculations and the experimental data ($d^* \sim 2$ μm, see Table 2) is observed at the activation energy of diffusion in the nonequilibrium grain boundaries $\sim 7.6$ $kT_m$ ($\sim 114$ kJ/mol). This value of activation energy is typical for the UFG metals [30].

**Conclusions**

1. The UFG austenitic steel 321L samples having good mechanical properties (the ultimate strength up to 1100 MPa) and good elongation to failure at room temperature were made by ECAP. The mass fraction of martensite after four ECAP cycles at 150 and 450 °C was ~5-6% and~15%, respectively. When heating the UFG steel 321L samples, nucleation of σ-phase nanoparticles was observed, which provides an increased thermal stability of the nonequilibrium UFG microstructure.

2. The dependencies of the elongation to failure on the testing temperature for the UFG steel 321L samples have a non-monotonous character with minima at 450 °C. Similar character of the dependence δ(T) was observed for coarse-grained steel. When increasing the number of ECAP cycles, a decrease in the elongation to failure of UFG steel 321L at 450 °C was observed. The result evidences indirectly the strain-induced formation of martensite in the UFG steel 321L to limit the elongation to failure of the steel when tensing in the temperature range from room temperature up to 450 °C.

3. The UFG steel 321L samples have good elongation to failure at 600-900 °C. The magnitudes of the strain rate sensitivity coefficient in the samples made by ECAP at 150 and 450 °C $m \sim$ 0.16-0.18 and 0.26-0.29, respectively. This allows suggesting the optimal regimes of superplasticity were not achieved in these temperature-rate deformation conditions. The values of maximum elongation to failure ($\delta_{max}$ = 250%) in the UFG steel 321L samples is limited by the cavitation of the samples. The formation of large pores at the non-metallic particles TiN and Ti(C,N) as well as of the micron-sized pores at the σ-phase particles was shown to be the main origin of the cavitation failure.

4. In certain temperature-rate conditions (600-900 °C, $3.3 \cdot 10^{-3} – 3.3 \cdot 10^{-3}$ s$^{-1}$), the superplastic deformation of the UFG steel 321L samples is controlled simultaneously by the grain boundary sliding and by the power law creep. The contribution of each process is determined, first of all, by the intensity of grain boundary migration as well as by the defect accumulation at the nonequilibrium grain boundaries in UFG steel in the superplasticity conditions.

**Acknowledgements**

The present study was supported by Russian Science Foundation (Grant No. 22-19-00238).

The TEM microstructure were carried out the equipment of Center Collective Use "Materials Science and Metallurgy" (National University of Science and Technology "MISIS") with the financial support of the Ministry of Science and Higher Education of the Russian Federation (Grant No. 075-15-2021-696).

**Author Contribution Statement**

**V.I. Kopylov** – Project administration, Investigation (obtained by UFG steel by ECAP, optimization of ECAP modes), Writing of manuscript, Analysis of experimental results, Funding acquisition. **M.Yu. Gryaznov** & **S.V. Shotin** – Investigation (mechanical tensile tests). **A.V. Nokhrin** – Investigation (XRD**),** Analysis of experimental results, Writing of manuscript, Data curation. **C.V. Likhnitskii** – Investigation (metallography, microhardness). **M.K. Chegurov** – Investigation (SEM, fractography). **V.N. Chuvil'deev** - Methodology, Analysis of experimental results, Writing - review & editing. **N.Yu. Tabachkova** - Investigation (TEM).

**Conflict of interest**. The authors declare that they have no conflict of interest.

**Data Availability**. The raw/processed data required to reproduce these findings cannot be shared at this time as the data also forms part of an ongoing study.